\documentclass[11pt,letterpaper]{article}
\usepackage{yuanbao-report}
\usepackage{amsmath,booktabs,array}
\usepackage{placeins,needspace}
\tcbuselibrary{breakable}
\newcommand{\model}{Yuanbao-Reranker}
\title{Yuanbao-Reranker}
\newcommand{\reportsubtitle}{Document-Grounded Synthesis and Preference Distillation for Text Reranking}
\hypersetup{pdftitle={Yuanbao-Reranker: Document-Grounded Synthesis and Preference Distillation for Text Reranking},pdfdisplaydoctitle=true}
\author{%
  {\bfseries Hongyuan Xu\textsuperscript{*},\enspace
  Dezhi Ye\textsuperscript{*\dag\ddag},\enspace
  Jiahao Cao,\enspace
  Jie Liu,\enspace
  Haijin Liang,\enspace
  Haibo Shi,\enspace
  Jin Ma}\\[3pt]
  {\color{reportmuted}Yuanbao Search, Tencent Inc.}
}
\githuburl{https://github.com/Yuanbao-Search/Yuanbao-Reranker}
\hfurl{https://huggingface.co/collections/Yuanbao-Search/yuanbao-reranker}
\date{}
\begin{document}
\maketitle
\begin{abstract}
High-precision text reranking is a critical component in modern search, recommendation, and agentic retrieval-augmented generation (RAG) systems. We present \model{}, a family of general-purpose dense rerankers spanning 0.6B, 4B, and 8B parameter scales built upon the Qwen3 backbone. To construct diverse and reliable ranking supervision at scale, we pair answer-first and query-type-guided document-grounded synthesis with a three-stage curation funnel that verifies positive support, eliminates semantic redundancy, and enforces balanced source-by-length quotas. For each curated pair, we screen real retrieval pools with a teacher score margin to filter out potential false negatives and sample Hard, Medium, and Moderate negative tiers, yielding 2.75M $1{:}3$ ranking groups. We train the models by jointly optimizing pointwise relevance cross-entropy and a bounded minimum-gap hinge objective over both positive--negative and negative--negative pairs. This formulation transfers fine-grained teacher preferences across candidate tiers without over-constraining well-separated predictions, supporting both cross-scale and same-scale distillation followed by layerwise adaptive model merging. Across five comprehensive benchmarks spanning Chinese (CMTEB-R), multilingual (MMTEB-R), English (BEIR), long-document (MLDR), and code (MTEB-Code) retrieval, \model{} consistently leads every parameter tier: the 0.6B variant reaches 70.96 Avg.\ to surpass 1B--4B active-parameter baselines, the 4B variant reaches 73.63 Avg.\ to outperform the 8B teacher (72.68), and the 8B variant achieves a state-of-the-art 74.21 Avg., with substantial gains generalizing across distinct first-stage retrievers.
\end{abstract}
\begin{figure}[!htb]
 \centering
 \vspace{-1.0mm}
 \includegraphics[width=0.94\linewidth]{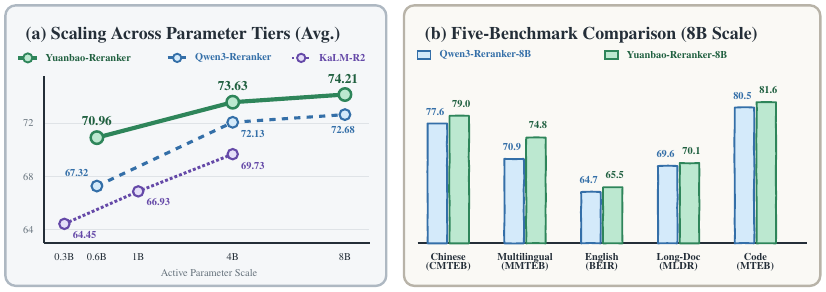}
 \vspace{-1.5mm}
 \caption{\textbf{Overview of \model{} performance across parameter scales on Qwen Top-100} (top-100 candidates per query retrieved by Qwen3-Embedding-0.6B). \textbf{(a)} Five-benchmark average score (Avg.) across active parameter scales. \textbf{(b)} Five-benchmark comparison at the 8B scale across Chinese, multilingual, English, long-document, and code retrieval.}
 \label{fig:teaser}
 \vspace{-2.0mm}
\end{figure}
\clearpage
\begingroup
\setlength{\parskip}{3.6pt plus 0.5pt minus 0.5pt}
\enlargethispage{1.5\baselineskip}
\section{Introduction}
\label{sec:introduction}
Text reranking models are a fundamental component of modern information retrieval, recommendation, and retrieval-augmented generation (RAG) systems. Operating as the precision stage after first-stage candidate generation, a cross-encoder reranker jointly encodes the query and each candidate passage to capture fine-grained contextual interactions~\citep{nogueira2019passage}. Recent rerankers built upon foundation language models have substantially advanced instruction following, multilingual understanding, and long-context modeling~\citep{zhang2025qwen3embedding,cai2025erank}. At the same time, practical deployments span diverse latency and accuracy budgets, motivating a unified model family covering compact models (0.6B and 4B) for high-throughput serving and a larger 8B model for peak ranking accuracy.

Despite recent progress, building a general-purpose reranker family that performs strongly across parameter scales remains challenging. Unlike coarse topical matching, reranking requires verifying whether a candidate passage satisfies specific query conditions, comparisons, or domain constraints. Existing training pipelines face three closely related limitations. First, public relevance benchmarks cover a narrow range of domains and query formulations, while unconstrained query synthesis with large language models~\citep{dai2022promptagator,jeronymo2023inparsv2} often omits qualifying conditions, leaks answers into the query text, or repeats simple factual lookups. Second, naive score-based data selection tends to over-select short encyclopedic passages relative to long or technical documents. In addition, top-ranked dense retrieval candidates frequently contain unlabelled relevant passages (\emph{false negatives})~\citep{qu2021rocketqa,cai2022hardnegatives} that introduce conflicting supervision if penalized, or narrow the intra-group difficulty spectrum if sampled only from the hardest edge. Third, standard pointwise fine-tuning on binary labels (\texttt{yes}/\texttt{no}) treats all negative candidates equally and ignores relative relevance gradations among competing distractors, whereas directly regressing raw teacher score differences can over-constrain the student on already well-separated pairs.

In this work, we introduce \model{}, a family of dense pointwise text rerankers spanning \textbf{0.6B, 4B, and 8B} parameter scales. To build a strong and general-purpose reranker family, we design an end-to-end data-construction and knowledge-transfer framework that progressively moves from grounded query synthesis to fine-grained groupwise preference learning (Figure~\ref{fig:pipeline}). To address query scarcity and ungrounded generation (Section~\ref{sec:data}), we combine \emph{answer-first basic synthesis}, which grounds each query in a complete answer proposition and its verbatim supporting span, with \emph{query-type-guided synthesis} structured around diverse information-seeking intents and iterative prompt auditing. To counter selection bias and false negatives, a three-stage curation pipeline (teacher verification, semantic deduplication, and source-by-length quota allocation) filters out partially supported pairs while preserving balanced coverage across domains and document lengths. For each curated pair, we screen real retrieval pools with a teacher score margin to exclude potential false negatives and sample three difficulty tiers (\emph{Hard}, \emph{Medium}, and \emph{Moderate}) per query, producing 2.75M $1{:}3$ ranking groups.

To transfer fine-grained relevance gradations without over-constraining student predictions (Section~\ref{sec:training}), we train \model{} jointly with pointwise relevance cross-entropy and a \emph{bounded minimum-gap hinge objective} over eligible positive--negative and negative--negative pairs within each group. Formulating capped teacher score margins as one-sided lower bounds transfers relative teacher preferences across candidate tiers while remaining compatible with pointwise label supervision, supporting both cross-scale distillation (to 0.6B and 4B) and same-scale distillation (at 8B), followed by layerwise \emph{Adaptive LERP} model merging. Across five benchmark panels covering Chinese (CMTEB-R), multilingual (MMTEB-R), English (BEIR), long-document (MLDR), and code (MTEB-Code) retrieval, \model{} establishes new state-of-the-art results within each parameter tier (Figure~\ref{fig:teaser}): \textbf{\model{}-0.6B} (70.96 Avg.) outperforms Qwen3-Reranker-0.6B ($+3.64$) and 1B--4B baselines such as KaLM-Large-R2 (69.73); \textbf{\model{}-4B} (73.63 Avg.) surpasses the 8B teacher (72.68); and \textbf{\model{}-8B} reaches 74.21 Avg. Beyond headline scores, controlled ablations and group-level diagnostics (Section~\ref{sec:analysis}) confirm that our bounded minimum-gap objective avoids the over-separation conflicts of squared-margin regression, while negative--negative contrasts substantially sharpen fine-grained ordering across competing negative tiers, driving pronounced gains on long-document retrieval. Evaluations with an independent first-stage embedding model further verify that these improvements generalize robustly across candidate distributions.

\par
\endgroup
\clearpage
\section{Data Recipe}
\label{sec:data}
Our data pipeline is designed to address three practical issues in reranker training data: limited query diversity in public benchmarks, domain and length imbalance when filtering synthetic pairs, and false negatives among retrieved candidates.

\begin{figure}[t!]
 \centering
 \includegraphics[width=\linewidth]{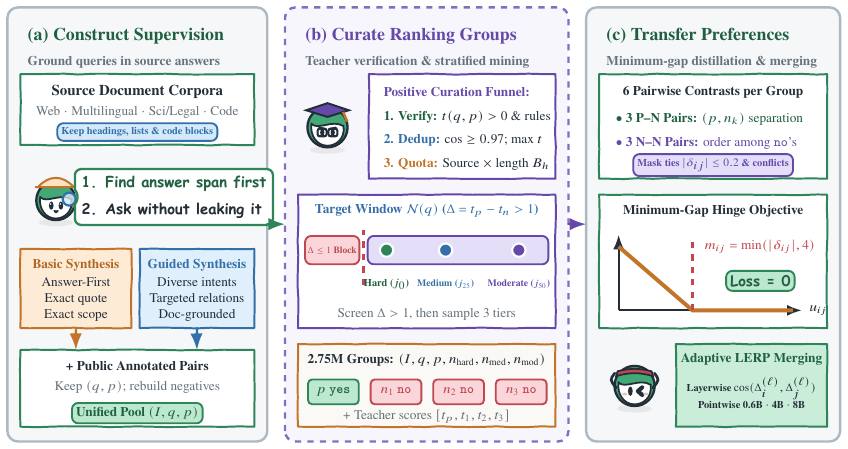}
 \caption{\textbf{Overview of the \model{} pipeline.} \textbf{(a) Construct Supervision:} Basic (answer-first) and query-type-guided synthesis generate query--positive pairs from source documents alongside public annotations. \textbf{(b) Curate Ranking Groups:} Teacher scores verify positive support and screen real retrieval pools, while embedding deduplication and source--length quotas balance coverage to assemble $1{:}3$ groups across Hard, Medium, and Moderate negative tiers ($j_0, j_{25}, j_{50}$). \textbf{(c) Transfer Preferences:} Binary labels supervise pointwise relevance via SFT, while bounded minimum-gap hinge distillation transfers pairwise teacher preferences into 0.6B, 4B, and 8B models.}
 \label{fig:pipeline}
\end{figure}

Figure~\ref{fig:pipeline} illustrates the four stages of our pipeline. First, we standardize public relevance annotations and unlabelled document collections into a common format that preserves document structure (Section~\ref{sec:taskdata}). Second, we expand query--positive pairs through \emph{answer-first basic synthesis} and \emph{query-type-guided synthesis}, refined with iterative prompt review (Section~\ref{sec:synthesis}). Third, we filter and balance the synthetic pool through teacher relevance verification, semantic deduplication, and source-by-length quota allocation (Section~\ref{sec:qpquality}). Finally, we retrieve candidate negatives from real corpora, exclude potential false negatives using teacher score margins, and sample \emph{Hard}, \emph{Medium}, and \emph{Moderate} negatives to build 2.75M $1{:}3$ ranking groups (Section~\ref{sec:negatives}).

\subsection{Data Collection and Preparation}
\label{sec:taskdata}
We combine public relevance benchmarks with unlabelled document collections across multiple languages and domains. Public annotation sources, including MS MARCO~\citep{bajaj2016msmarco}, Natural Questions~\citep{kwiatkowski2019naturalquestions}, HotpotQA~\citep{yang2018hotpotqa}, DuReader~\citep{he2018dureader,qiu2022dureaderretrieval}, T$^2$-Ranking~\citep{xie2023t2ranking}, and MIRACL~\citep{zhang2023miracl}, provide verified query--positive pairs. Our document collections span encyclopedic articles, academic papers, legal and medical texts, technical manuals, and code repositories, supplying source material for query synthesis (listed in Appendix~\ref{app:data-sources}).

For document collections, source adapters extract the main text or structured content fields, remove navigation elements and page templates, and normalize text formatting. We preserve structural elements (headings, paragraphs, lists, tables, and code blocks) so that query synthesis can use document organization and context. Each prepared document retains its language and source metadata.

For public annotations, we convert queries, positive passages, and task instructions into a shared format while keeping all known positive links. Even when a dataset provides pre-sampled negatives, we rebuild its negative candidates through our unified retrieval and teacher-screening procedure (Section~\ref{sec:negatives}). Rebuilding negatives consistently across public and synthetic pairs reduces dataset-specific pooling bias and removes unjudged relevant documents that would otherwise act as false negatives.

\Needspace{8\baselineskip}
\subsection{Document-Grounded Synthesis}
\label{sec:synthesis}
While unlabelled corpora cover diverse topics, naive question generation often produces queries that are unsuitable for reranker training: generated questions may introduce unsupported assumptions, leak the answer inside the query, or repeat simple factual lookups. To generate reliable training pairs~\citep{dai2022promptagator,jeronymo2023inparsv2,zhang2025qwen3embedding}, we combine \emph{basic synthesis} with \emph{query-type-guided synthesis}, supported by iterative prompt review.

\paragraph{Basic Synthesis: Establish why the document is a positive.}
A query that shares keywords with a document may still ask for information the document does not provide. Our \emph{answer-first} procedure reverses standard prompting: the generator first reads the full document, identifies a complete answer proposition, and locates its exact supporting quote before writing a natural, standalone query.

Authored queries must keep key conditions, such as entities, time ranges, jurisdictions, and operational constraints, while leaving the answer inside the document. The authoring rules disallow generic summary requests (e.g., ``What is this article about?''), metadata lookups, and paraphrases of questions already present in the source text. The positive document is kept unchanged, and noisy or incomplete documents are filtered out.

\paragraph{Query-Type-Guided Synthesis: Broaden information needs.}
Without guidance, language models tend to ask about prominent declarative facts, whereas practical retrieval tasks span a much broader range of query--document relations. Examples include, among others, troubleshooting software or systems issues against technical documentation, consulting applicable regulations or clinical guidelines for practical scenarios, locating supporting or contrasting evidence for analytical claims, and exploring scholarly literature or community knowledge.

To cover such diverse retrieval patterns, we organize representative information-seeking intents into structured authoring instructions and instantiate them against public source documents. Here, the query-type specification guides \emph{how} a request is framed, while the source document determines \emph{what} concrete facts, arguments, or conditions are supported. Both synthesis modes follow the same grounding rules and enter the same curation pipeline.

\paragraph{Iterative Prompt Refinement: Generalize from recurring errors.}
Before large-scale generation, we run an iterative generation-and-audit cycle in which candidate pairs are reviewed against their full source documents. Recurring failure patterns are summarized into general authoring rules, added to the prompt, and tested on a \emph{new batch of documents} until the review pass rate reaches the target level. For example, observing dropped conditional clauses motivated our scope-preservation rule, while testing on new batches ensures that prompt updates generalize beyond the reviewed samples (Appendix~\ref{app:synthesis-example}).

\paragraph{Design Principle: Why Positive Synthesis + Real Negative Mining?}
In preliminary experiments, neither query-to-negative-document synthesis nor document-to-negative-query synthesis improved ranking performance. Generating synthetic negative documents faces a trade-off: moving too far from the query produces easy negatives, while staying close risks keeping relevant answers. In addition, LLM-generated negatives often contain stylistic patterns that models can exploit as shortcuts~\citep{zhang2026causalneg}. Conversely, generating a negative query for a fixed document creates only a single negative link without a paired positive passage or comparison group. Although negative-query synthesis is useful in multimodal reranking~\citep{wasserman2025docrerank}, our group-based training recipe works best by \textbf{synthesizing grounded positive pairs and mining competing negatives from real corpora}.

\subsection{Quality and Coverage Control}
\label{sec:qpquality}
We refine the raw synthetic pool $\mathcal D_{\mathrm{syn}}$ through a three-stage curation funnel ($\mathcal D_{\mathrm{syn}} \rightarrow \mathcal D_{\mathrm{qual}} \rightarrow \mathcal D_{\mathrm{dedup}} \rightarrow \mathcal D_{\mathrm{selected}}$) to remove partially supported pairs, near-duplicate queries, and source imbalance.

\begin{figure}[t!]
 \centering
 \includegraphics[width=\linewidth]{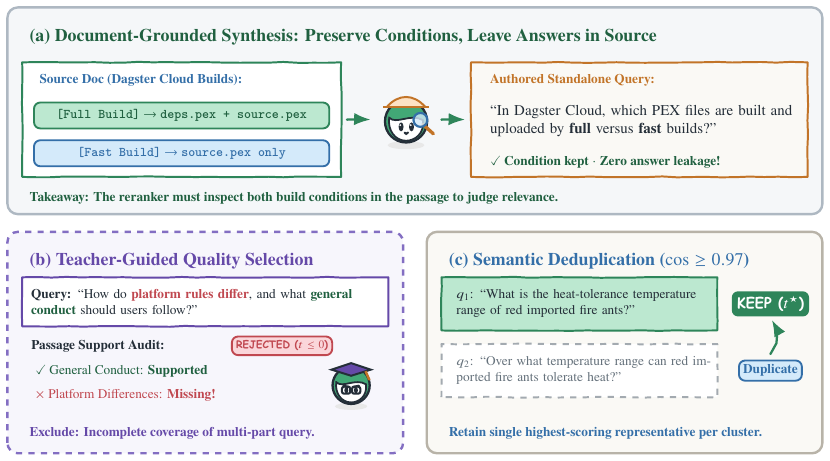}
 \caption{\textbf{Examples of synthetic-data construction.} (a) A generated query preserves a full-versus-fast build distinction while leaving the answer in the source. (b) The passage covers general conduct but lacks a requested platform comparison; teacher-guided selection excludes the pair. (c) Two supported pairs express the same information need, and deduplication retains one representative. Examples are translated and condensed for clarity.}
 \label{fig:data-case-studies}
\end{figure}

\Needspace{8\baselineskip}
\paragraph{Stage 1 (Teacher-Guided Quality Selection): Verify the positive relationship.}
To filter out pairs where the query is only partially answered by the source passage, we score each candidate pair $(q,p)\in\mathcal D_{\mathrm{syn}}$ with Qwen3-Reranker-8B~\citep{zhang2025qwen3embedding} under task instruction $I$:
\begin{equation}
 t(q,p)=z_T^{\mathrm{yes}}(I,q,p)-z_T^{\mathrm{no}}(I,q,p),
 \label{eq:curation-teacher-score}
\end{equation}
where $z_T^{\mathrm{yes}}$ and $z_T^{\mathrm{no}}$ are the teacher logits for the relevance decision tokens. We require $t(q,p)>\tau_{\mathrm{pos}}$ ($\tau_{\mathrm{pos}}=0$) together with rule checks $H(q,p)=1$ that exclude title copies and unanchored relative-time expressions:
\begin{equation}
 \mathcal D_{\mathrm{qual}}=
 \left\{(q,p)\in\mathcal D_{\mathrm{syn}}:
 t(q,p)>\tau_{\mathrm{pos}},\ H(q,p)=1\right\},
 \qquad \tau_{\mathrm{pos}}=0.
 \label{eq:positive-quality-selection}
\end{equation}
As shown in Figure~\ref{fig:data-case-studies}(b), teacher verification removes pairs where the passage covers general background conduct but lacks a requested comparison.

\paragraph{Stage 2 (Semantic Deduplication): Consolidate repeated information needs.}
Similar documents can yield near-duplicate queries. We encode queries in $\mathcal D_{\mathrm{qual}}$ using Qwen3-Embedding-0.6B into $\ell_2$-normalized vectors $\mathbf e_i=E(q_i)/\|E(q_i)\|_2$ and link same-language queries with cosine similarity at least $\tau_{\mathrm{dup}}=0.97$:
\begin{equation}
 (i,j)\in\mathcal E
 \iff \operatorname{lang}(q_i)=\operatorname{lang}(q_j)
 \ \land\ \mathbf e_i^\top\mathbf e_j\ge\tau_{\mathrm{dup}},
 \qquad \tau_{\mathrm{dup}}=0.97.
 \label{eq:query-dedup-edges}
\end{equation}
Connected components in $\mathcal E$ define query clusters. From each cluster $C$, we keep the single pair with the highest teacher score:
\begin{equation}
 (q_C^\star,p_C^\star)
 =\underset{(q,p)\in C\cap\mathcal D_{\mathrm{qual}}}{\arg\max}\ t(q,p),
 \label{eq:query-cluster-representative}
\end{equation}
and remove exact query or document duplicates against existing data to form $\mathcal D_{\mathrm{dedup}}$ (Figure~\ref{fig:data-case-studies}(c)).

\paragraph{Stage 3 (Coverage-Aware Selection): Prevent high-scoring strata from dominating.}
Teacher scores vary across domains and document lengths: short encyclopedic passages typically receive higher raw scores than long technical or legal documents. Global top-$B$ selection would therefore over-select short, easy sources.

To separate quality ranking from source distribution, we set budgets at the domain--language level and divide each source into \emph{short}, \emph{medium}, and \emph{long} character-length bands using its $1/3$ and $2/3$ length quantiles. Within a domain--language group with total budget $B$, let $\mathcal D_h \subseteq \mathcal D_{\mathrm{dedup}}$ denote the deduplicated candidates in source--length stratum $h$, with $n_h=|\mathcal D_h|$ and $N=\sum_h n_h$. We allocate stratum quotas proportionally:
\begin{equation}
 \widetilde B_h=B\frac{n_h}{N}.
 \label{eq:coverage-allocation}
\end{equation}
After assigning $\lfloor\widetilde B_h\rfloor$ places and distributing remaining places by largest fractional remainders to obtain integer quotas $B_h$ ($\sum_h B_h=B$), we select the highest-scoring pairs \emph{within each stratum}:
\begin{equation}
 \mathcal D_{\mathrm{selected}}^{\mathrm{group}}
 =\bigcup_h\operatorname{Top}_{B_h}
 \bigl(\mathcal D_h;\ t(q,p)\bigr).
 \label{eq:coverage-selection}
\end{equation}
\textbf{Stratum quotas control coverage across sources and lengths, while teacher scores select the best examples within each stratum} (Appendix~\ref{app:selection-settings}).

\Needspace{8\baselineskip}
\subsection{Negative Mining and Group Construction}
\label{sec:negatives}
Effective reranker training requires negative documents that are topically related to the query but genuinely non-relevant. Because top-ranked retrieval candidates often include unlabelled positives~\citep{qu2021rocketqa,cai2022hardnegatives}, we separate candidate retrieval, teacher-guided false-negative filtering, and difficulty-stratified sampling.

\paragraph{Candidate Retrieval: Mine realistic competitors.}
For each query--positive pair (both public and synthetic), we retrieve candidate passages from source-compatible corpora with language routing, excluding known positive IDs and duplicate texts.

\paragraph{Teacher-Guided Screening: Exclude potential false negatives.}
Using the 8B teacher score $t(q,d)$ from Equation~\ref{eq:curation-teacher-score}, we compute the score gap between the positive $p$ and each retrieved candidate $n\in\mathcal C(q)$:
\begin{equation}
 \begin{aligned}
 \Delta(q,n)&=t(q,p)-t(q,n),\\
 \mathcal N(q)&=\{n\in\mathcal C(q):\Delta(q,n)>\tau_{\mathrm{neg}}\},
 \qquad \tau_{\mathrm{neg}}=1.
 \end{aligned}
 \label{eq:negative-selection}
\end{equation}
Candidates with $\Delta(q,n)\le 1$ are excluded as potential false negatives, reducing the risk of penalizing unlabelled relevant documents.

\paragraph{Difficulty-Stratified Selection: Sample Hard, Medium, and Moderate negatives.}
We sort the eligible negative set $\mathcal N(q)$ by ascending teacher gap $\Delta(q,n)$, so smaller indices correspond to harder competitors closer to the positive. Rather than selecting all three negatives from the hardest edge, which would narrow the difficulty range and reduce score differences among negatives, we sample three distinct documents at the \textbf{top-ranked (\emph{Hard}, $j_0$)}, \textbf{first-quartile (\emph{Medium}, $j_{25}$)}, and \textbf{median (\emph{Moderate}, $j_{50}$)} rank positions of $\mathcal N(q)$, while skipping the easier half beyond the median.

\begin{takeawaybox}
\textbf{Group-Construction Principle:} Teacher screening ($\Delta > 1$) ensures that negative candidates are safe to contrast, while quantile sampling across \emph{Hard} ($j_0$), \emph{Medium} ($j_{25}$), and \emph{Moderate} ($j_{50}$) rank positions provides clear relevance gradations for ranking distillation.
\end{takeawaybox}

Each query--positive pair and its three stratified negatives form a $1{:}3$ ranking group:
\begin{equation}
 g=(I,q,p,n_{\mathrm{hard}},n_{\mathrm{med}},n_{\mathrm{mod}}),
 \label{eq:training-group}
\end{equation}
paired with binary labels $[\texttt{yes},\texttt{no},\texttt{no},\texttt{no}]$ and detached teacher scores $[t_p,t_1,t_2,t_3]$. This group structure supports four pointwise SFT decisions and up to six pairwise comparisons (three positive--negative and three negative--negative) for distillation. Complete groups are shuffled and batched together during training, totaling approximately 2.75M ranking groups (discrete index formulas in Appendix~\ref{app:negative-settings}).

\section{Training Recipe}
\label{sec:training}
We adapt dense Qwen3 language models to reranking through supervised fine-tuning and teacher-guided ranking distillation. Both objectives operate on the query-centered groups constructed in Section~\ref{sec:data}: binary relevance labels supervise individual query--document decisions, while teacher scores provide relative preferences within each group. Throughout training and deployment, the underlying model architecture and pointwise inference interface remain unchanged.

\subsection{Supervised Fine-Tuning}
Following the input formulation of Qwen3-Reranker~\citep{zhang2025qwen3embedding}, \model{} jointly encodes a task instruction, query, and candidate document to predict a binary relevance token (\texttt{yes} or \texttt{no}). We initialize the 0.6B, 4B, and 8B models from the corresponding instruction-following Qwen3 backbones and fine-tune them with full-vocabulary token cross-entropy over the response position, masking all prompt tokens from the loss. Positive documents receive the target \texttt{yes}, and negatives receive \texttt{no}.

\begin{tcolorbox}[enhanced,colback=reportpale,colframe=reportaccent!45,boxrule=0.5pt,arc=1.2mm,left=3.5mm,right=3.5mm,top=2mm,bottom=2mm,title={Reranking Prompt},colbacktitle=reportaccentlight,coltitle=reportaccent,fonttitle=\sffamily\bfseries,fontupper=\fontsize{9.4}{11.6}\selectfont]
\textbf{System}\par
Judge whether the Document meets the requirements based on the Query and the Instruct provided. Note that the answer can only be ``yes'' or ``no''.
\par\smallskip
\textbf{User}\par
\texttt{<Instruct>: \{task\_instruction\}}\par
\texttt{<Query>: \{query\}}\par
\texttt{<Document>: \{document\}}
\par\smallskip
\textbf{Assistant Target}\par
\texttt{\{yes\_or\_no\}}
\end{tcolorbox}
The task instruction specifies the retrieval objective. The template above omits chat delimiters and the empty \texttt{<think>\ldots</think>} block preceding the non-thinking response.

For candidate document $d_i$, we define the relevance score at the decision position as the logit margin between the two decision tokens:
\begin{equation}
 s_i=s_\theta(I,q,d_i)=z_i^{\mathrm{yes}}-z_i^{\mathrm{no}},
 \label{eq:studentscore}
\end{equation}
where $z_i^{\mathrm{yes}}$ and $z_i^{\mathrm{no}}$ denote the corresponding token logits and $I$ is the task instruction. Larger scores indicate higher relevance; applying a two-way softmax over the decision tokens yields $r_i=\sigma(s_i)$, which is strictly monotonic in $s_i$ and preserves the induced document ordering. Because each candidate is scored independently, we augment this pointwise objective during training with teacher-guided pairwise preferences, as detailed next.

\subsection{Ranking Distillation}
\label{sec:hinge}
Teacher-dependent pairwise margins are also used in Querit-Reranker~\citep{zhong2026querit}, whose published objective scales the target separation linearly with the teacher-score gap. We adapt this hinge formulation to our label-supervised ranking groups through three complementary choices: \textbf{capped margins} limit the separation demanded by large teacher logit gaps; \textbf{selective comparisons} omit near-ties and annotation-conflicting groups from distillation; and \textbf{joint SFT supervision} retains the original relevance targets alongside the relative ranking objective.

Qwen3-Reranker-8B~\citep{zhang2025qwen3embedding} supplies teacher scores using the same yes-minus-no logit-margin formulation. Let $t_i$ denote the detached teacher score for document $i$. For each unordered document pair $i<j$ within a ranking group, we define the teacher score difference, the teacher-oriented student gap, and the capped target margin as
\begin{equation}
 \delta_{ij}=t_i-t_j,\qquad
 u_{ij}=\operatorname{sign}(\delta_{ij})(s_i-s_j),\qquad
 m_{ij}=\min\bigl(|\delta_{ij}|,m_{\max}\bigr).
 \label{eq:margin}
\end{equation}
The teacher provides both a directional preference and a target separation magnitude, bounded above by $m_{\max}=4$. Below the cap, the target retains the teacher's gap magnitude; above it, the required separation no longer grows with the logit difference. This limits the distance the student must reproduce for already well-separated teacher predictions. Each $1{:}3$ group yields six candidate comparisons: three positive--negative pairs and three negative--negative pairs.

Near-equal teacher scores---particularly between adjacent negative tiers---offer little separation to transfer, while a teacher preference that contradicts a human-annotated positive in public benchmarks would oppose the supervised target. We therefore mask near-tied pairs and annotation-conflicting groups from ranking distillation while retaining their SFT supervision. Let $E_g=\mathbf{1}\!\left[t_p \ge \max_k t_{n_k}\right]$ indicate group-level teacher consistency with the positive label. With tie threshold $\epsilon=0.2$, the eligible-pair indicator and active pair count are
\begin{equation}
 a_{ij}=E_g\,\mathbf{1}\!\left[|\delta_{ij}|>\epsilon\right],
 \qquad K_g=\sum_{i<j}a_{ij}.
\end{equation}
We optimize the average hinge penalty over eligible pairs jointly with the supervised fine-tuning loss:
\begin{align}
 \mathcal{L}_{\mathrm{rank}}
 &=\frac{1}{G}\sum_{g=1}^{G}
 \frac{\sum_{i<j}a_{ij}\,[m_{ij}-u_{ij}]_+}{\max(K_g,1)},
 \label{eq:rankloss}\\
 \mathcal{L}
 &=\mathcal{L}_{\mathrm{SFT}}+\lambda\mathcal{L}_{\mathrm{rank}},
 \label{eq:totalloss}
\end{align}
where $[x]_+=\max(x,0)$, $G$ is the number of groups in the batch, and $\lambda$ balances the distillation term. If a group has no eligible pairs ($K_g=0$), its ranking loss vanishes while its binary labels continue to contribute to $\mathcal{L}_{\mathrm{SFT}}$, preserving the positive supervision whenever the teacher disagrees.

The two objectives govern complementary aspects of the student's scoring function. Whereas $\mathcal{L}_{\mathrm{SFT}}$ supervises each document's relevance decision, $\mathcal{L}_{\mathrm{rank}}$ is invariant to uniform score shifts within a group and shapes relative separations: positive--negative pairs sharpen the decision boundary, and negative--negative pairs expose fine-grained relevance gradations among non-relevant candidates that binary labels cannot express. These pairwise constraints couple candidate scores solely during training, leaving pointwise inference untouched.

\begin{figure}[t!]
 \centering
 \includegraphics[width=\linewidth]{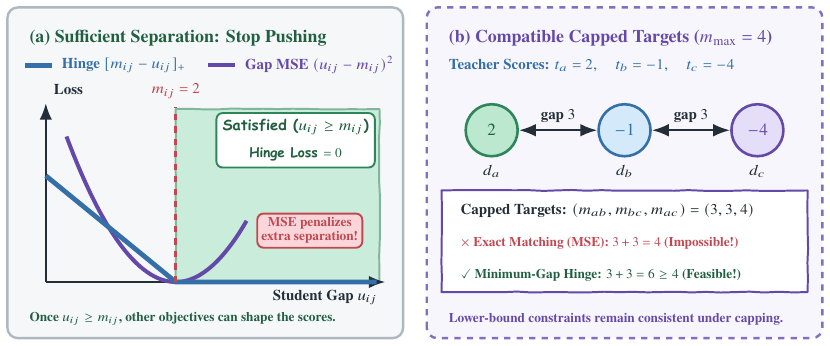}
 \caption{\textbf{Minimum-gap distillation.} \textbf{(a)} Hinge loss penalizes insufficient separation ($u_{ij}<m_{ij}$) and vanishes once the target margin is met, whereas squared gap matching penalizes extra separation. \textbf{(b)} Teacher scores $(2,-1,-4)$ yield capped targets $(3,3,4)$ under $m_{\max}=4$, which violate exact path additivity ($3+3\neq 4$) but remain mutually consistent as lower bounds ($3+3=6\ge 4$).}
 \label{fig:marginshape}
\end{figure}

\paragraph{Minimum separation rather than exact gap matching.}
Relative-score distillation commonly treats teacher margins as exact regression targets~\citep{hofstatter2020marginmse}. In contrast, our hinge formulation treats each target margin $m_{ij}$ as a lower bound: the pairwise penalty drops to zero once the student's oriented gap $u_{ij}$ reaches $m_{ij}$. Consequently, $\mathcal{L}_{\mathrm{SFT}}$ and other active pairs can further adjust individual scores without incurring a penalty for exceeding the target separation on an already-satisfied pair.

This lower-bound view also resolves the geometric inconsistency introduced by margin capping. True scalar score differences are path-additive, whereas independently capped pairwise gaps need not be. For instance, teacher scores $(2,-1,-4)$ induce raw gaps of $3$, $3$, and $6$, which become $(3,3,4)$ under $m_{\max}=4$. No three scalar student scores can match $(3,3,4)$ with equality, yet the teacher scores themselves trivially satisfy all three minimum-gap inequalities (Figure~\ref{fig:marginshape}). More generally, the teacher score vector always lies in the feasible region of the retained lower-bound constraints within any valid group (see Appendix~\ref{app:margin-consistency}). This guarantees that the distillation targets remain mutually consistent without requiring a compact student to replicate the teacher's exact unbounded logit geometry.

\subsection{Training Configuration}
We fine-tune all model parameters for one epoch over the 2.75M-group mixture with a maximum sequence length of 4,096 tokens. Each global batch comprises 64 ranking groups (256 query--document examples). Optimization uses AdamW with a peak learning rate of $10^{-5}$, cosine learning-rate decay, a 2\% linear warmup, and a distillation weight of $\lambda=0.1$. Each ranking group is kept intact on a data-parallel worker so eligible pairwise comparisons are computed within the same group, with teacher scores and margin arithmetic computed in FP32 alongside BF16 model computation. Our matched SFT-only comparison uses the same initialization, training data, and optimization schedule with $\lambda=0$. Complete hyperparameters are listed in Appendix~\ref{app:train}.

\Needspace{12\baselineskip}
\subsection{Model Merging}
\label{sec:merging}
\label{sec:adaptive-merging}
Following training, we integrate models that capture complementary ranking signals through parameter interpolation, balancing performance across diverse retrieval distributions while preserving a single-model footprint and identical inference cost~\citep{wortsman2021robust,zhang2025qwen3embedding}. In this setup, we designate a primary model as the anchor and incorporate parameter updates from auxiliary models via \emph{Adaptive LERP}, a layerwise scheme that modulates auxiliary weights according to the directional consistency of their parameter updates. Because fine-tuned checkpoints need not align uniformly across network depth, layers where auxiliary updates point in a consistent direction relative to the primary model receive a larger auxiliary weight, whereas layers with divergent updates retain a stronger anchor weight from the primary model.

Concretely, let $\theta_0$ denote the primary model and $\mathcal{S}$ ($|\mathcal{S}|\ge 2$) index the set of auxiliary models obtained from complementary training runs initialized from the same Qwen3 backbone (varying data-mixture weights and training configurations). For each Transformer layer $\ell$, we flatten and concatenate all parameter tensors within that layer into a single vector and define the layerwise update of auxiliary model $i \in \mathcal{S}$ relative to the primary model as $\Delta_i^{(\ell)}=\theta_i^{(\ell)}-\theta_0^{(\ell)}$. We quantify the directional agreement of model $i$ with the remaining auxiliary models via the mean pairwise cosine similarity:
\begin{equation}
 c_i^{(\ell)}=\frac{1}{|\mathcal{S}|-1}
 \sum_{\substack{j\in\mathcal{S}\\j\ne i}}
 \cos\!\left(\Delta_i^{(\ell)},\Delta_j^{(\ell)}\right).
 \label{eq:merge-agreement}
\end{equation}
Using this agreement score, we assign the layerwise interpolation weight for each auxiliary model $i \in \mathcal{S}$ and the residual weight for the primary model as
\begin{equation}
 w_i^{(\ell)}=\frac{0.5}{|\mathcal{S}|}
 \left(0.25+0.75\max(0,c_i^{(\ell)})\right),
 \qquad w_0^{(\ell)}=1-\sum_{i\in\mathcal{S}}w_i^{(\ell)}.
 \label{eq:merge-weights}
\end{equation}
The merged parameters for layer $\ell$ are then obtained by the convex combination
\begin{equation}
 \theta_{\mathrm{adaptive}}^{(\ell)}
 =w_0^{(\ell)}\theta_0^{(\ell)}
 +\sum_{i\in\mathcal{S}}w_i^{(\ell)}\theta_i^{(\ell)}.
 \label{eq:adaptive-merge}
\end{equation}
By construction, $w_0^{(\ell)} \in [0.5, 0.875]$, ensuring that the primary model retains at least half of the total weight in every layer while selectively integrating consensual auxiliary updates. The resulting merged checkpoint is deployed as a single standard model with zero additional runtime overhead.

\Needspace{8\baselineskip}
\section{Experiments}
\label{sec:evaluation}
\subsection{Experimental Setup}
To thoroughly assess reranking capabilities across languages, domains, and document lengths, we evaluate all models across five complementary benchmark panels (full task lists and splits in Appendix~\ref{app:eval}):
\begin{itemize}
 \item \textbf{Chinese Retrieval (CMTEB-R):} Evaluates Chinese web search, e-commerce product matching, biomedical QA, and video/news passage retrieval across 8 tasks from the C-MTEB benchmark~\citep{xiao2023cpack}.
 \item \textbf{Multilingual \& Multitask Retrieval (MMTEB-R):} Evaluates cross-lingual and multilingual semantic matching across 16 tasks (49 evaluation units) spanning 18+ languages, legal statutes, dialogue tables, and temporal/spatial reasoning~\citep{enevoldsen2025mmteb}.
 \item \textbf{English Retrieval (BEIR):} Evaluates broad multi-domain generalization across 15 diverse English retrieval tasks covering fact verification, scientific citations, financial QA, multi-hop reasoning, and argument retrieval~\citep{thakur2021beir}.
 \item \textbf{Long-Document Retrieval (MLDR):} Evaluates fine-grained evidence localization within long documents (up to 8,192 tokens) across 13 typologically diverse languages~\citep{chen2024m3}.
 \item \textbf{Code \& Technical Retrieval (MTEB-Code):} Evaluates natural-language-to-code, code-to-code, SQL synthesis, and code-edit retrieval across 12 programming tasks (39 language subsets), including tasks from CoIR~\citep{enevoldsen2025mmteb,li2025coir}.
\end{itemize}

For first-stage retrieval, we use \textbf{Qwen3-Embedding-0.6B} to retrieve the top-100 candidate documents per query across all tasks, and refer to this shared candidate pool as \textbf{Qwen Top-100}. All evaluated rerankers score and reorder the exact same 100 retrieved candidates per query, holding first-stage recall fixed. We report nDCG@10 ($\times 100$) following each panel's standard task and language aggregation, and summarize overall performance by the equal-weight average (\textbf{Avg.}) across all five panels.

\subsection{Main Results}
\label{sec:mainresults}
Table~\ref{tab:main} compares the \model{} family (0.6B, 4B, and 8B) against open-weights rerankers grouped by active parameter scale on the shared Qwen Top-100 candidate pool.

\begin{table}[!htb]
\centering
\small
\setlength{\tabcolsep}{7.2pt}
\begin{tabularx}{\linewidth}{@{}Xlcccccc@{}}
\toprule
\rowcolor{tablehead}
Model & Params (Act.\ / Tot.) & CMTEB-R & MMTEB-R & BEIR & MLDR & Code & Avg. \\
\midrule
\multicolumn{8}{@{}l}{\emph{\textbf{Sub-1B Compact Models ($<1$B Active Parameters)}}} \\
mxbai-rerank-base-v2 & 0.5B & 69.77 & 59.04 & 58.22 & 42.99 & 68.01 & 59.61 \\
bge-reranker-v2-m3 & 0.57B & 72.14 & 58.27 & 55.48 & 59.41 & 41.34 & 57.33 \\
Querit-Reranker-A0.4B & 0.43B / 4.92B & 70.44 & 61.42 & 57.78 & 56.55 & 65.78 & 62.39 \\
KaLM-Reranker-Nano-R2 & 0.27B / 0.79B & 70.00 & 62.26 & 58.47 & 66.13 & 65.38 & 64.45 \\
jina-reranker-v3.5 & 0.6B & 73.96 & 66.71 & \textbf{61.32} & 65.01 & 73.90 & 68.18 \\
Qwen3-Reranker-0.6B & 0.6B & 71.43 & 65.61 & 59.61 & 66.89 & 73.03 & 67.32 \\
\rowcolor{tableresult}
\textbf{Yuanbao-Reranker-0.6B} & \textbf{0.6B} & \textbf{75.60} & \textbf{70.49} & 61.30 & \textbf{68.94} & \textbf{78.44} & \textbf{70.96} \\
\midrule
\multicolumn{8}{@{}l}{\emph{\textbf{Mid-Sized Models (1B--4B Active Parameters)}}} \\
KaLM-Reranker-Small-R2 & 1B / 2.12B & 73.18 & 64.17 & 60.90 & 67.29 & 69.09 & 66.93 \\
mxbai-rerank-large-v2 & 1.54B & 72.09 & 64.74 & 60.23 & 54.25 & 72.78 & 64.82 \\
bge-reranker-v2-gemma & 2.51B & 72.35 & 65.89 & 60.29 & 47.30 & 64.26 & 62.02 \\
Querit-Reranker-Dense-4B & 4B & 73.02 & 66.55 & 59.87 & 49.98 & 73.76 & 64.64 \\
KaLM-Reranker-Large-R2 & 4B / 7.51B & 75.58 & 66.86 & 62.72 & 69.72 & 73.75 & 69.73 \\
Qwen3-Reranker-4B & 4B & 76.05 & 71.29 & 63.66 & 69.35 & 80.33 & 72.13 \\
\rowcolor{tableresult}
\textbf{Yuanbao-Reranker-4B} & \textbf{4B} & \textbf{78.31} & \textbf{74.08} & \textbf{64.74} & \textbf{70.04} & \textbf{80.98} & \textbf{73.63} \\
\midrule
\multicolumn{8}{@{}l}{\emph{\textbf{Large Models (8B Parameters)}}} \\
Qwen3-Reranker-8B & 8B & 77.61 & 70.90 & 64.74 & 69.63 & 80.50 & 72.68 \\
\rowcolor{tableresult}
\textbf{Yuanbao-Reranker-8B} & \textbf{8B} & \textbf{78.99} & \textbf{74.84} & \textbf{65.51} & \textbf{70.12} & \textbf{81.58} & \textbf{74.21} \\
\bottomrule
\end{tabularx}
\caption{Reranking performance (nDCG@10 $\times 100$) on the shared \textbf{Qwen Top-100} candidate pool (top-100 candidates per query retrieved by Qwen3-Embedding-0.6B). Code denotes MTEB-Code, and \textbf{Avg.}\ is the equal-weight average across all five panels. For architectures that decouple active inference parameters from total checkpoint parameters (Querit-A0.4B~\citep{zhong2026querit} and KaLM-R2~\citep{zhao2026kalm}), both values are listed as \texttt{Active / Total}. Bold marks the best score within each tier.}
\label{tab:main}
\end{table}

\paragraph{Sub-1B Compact Models.}
\model{}-0.6B achieves \textbf{70.96 Avg.}, outperforming jina-reranker-v3.5 (68.18) by 2.78 points, Qwen3-Reranker-0.6B (67.32) by 3.64 points, and KaLM-Reranker-Nano-R2 (64.45) by 6.51 points. Notably, our 0.6B model outperforms all 1B--4B baselines outside the Qwen3 family, including KaLM-Reranker-Large-R2 (4B active / 7.51B total parameters, 69.73) and Querit-Reranker-Dense-4B (64.64).

\paragraph{Mid-Sized Models.}
\model{}-4B reaches \textbf{73.63 Avg.}, improving over Qwen3-Reranker-4B (72.13) by 1.50 points, KaLM-Reranker-Large-R2 (69.73) by 3.90 points, and surpassing the Qwen3-Reranker-8B teacher (72.68) by 0.95 points, with consistent improvements across all five evaluation panels.

\paragraph{Large Models.}
Training the 8B backbone on our curated 2.75M ranking groups via same-scale distillation and model merging yields \textbf{74.21 Avg.}, outperforming Qwen3-Reranker-8B by 1.53 points and achieving the highest score on every individual panel. Because document-grounded synthesis expands verified positive coverage while our bounded hinge objective ($\mathcal{L}_{\mathrm{SFT}}+\lambda\mathcal{L}_{\mathrm{rank}}$) enforces only local lower-bound separations ($m_{\max}=4$) without imitating uncalibrated pointwise teacher errors, both the 4B and 8B models surpass the 8B teacher.

\section{Analysis}
\label{sec:analysis}
To isolate the contribution of each methodological component without confounding effects from model merging, we conduct controlled ablations in Sections~\ref{sec:ablation-data}--\ref{sec:ablation-distill} using \textbf{0.6B Raw} checkpoints\footnote{Throughout Section~\ref{sec:analysis}, \textbf{Raw} denotes the checkpoint obtained directly after training (SFT + ranking distillation) prior to the model merging step in Section~\ref{sec:merging}.} evaluated on the shared Qwen Top-100 candidate pool (top-100 candidates per query retrieved by Qwen3-Embedding-0.6B). Section~\ref{sec:interpolation-study} then examines the effect of model merging across scales, and Section~\ref{sec:cross-retriever} evaluates generalization to a second first-stage retriever.

\subsection{Effect of Document-Grounded Synthesis and Negative Mining}
\label{sec:ablation-data}
\paragraph{Contribution of Curated Synthetic Supervision.}
We first evaluate the impact of document-grounded synthetic supervision by comparing a 0.6B Raw model trained solely on public annotated groups (already using our screened stratified $1{:}3$ negatives and Full Hinge distillation) against the full 2.75M mixture. As shown in the top block of Table~\ref{tab:ablation-synthesis-neg}, adding curated synthetic data raises the five-panel average from 68.77 to \textbf{70.43} ($+1.66$). Whereas public annotations already cover short Chinese and English web passages well (CMTEB-R, $+0.42$; BEIR, $+0.41$), synthetic supervision delivers pronounced gains where public data is scarce: long-document (MLDR, $+3.82$, from 64.52 to 68.34), multilingual and multitask (MMTEB-R, $+2.37$, from 67.53 to 69.90), and code retrieval (Code, $+1.27$, from 76.26 to 77.53). These results confirm that our synthesis and quota allocation remedy the length and domain gaps of public benchmarks, while Figure~\ref{fig:data-case-studies} illustrates how teacher verification and deduplication filter out unsupported or redundant pairs.

\paragraph{Impact of Group Size and Stratified Negative Mining.}
We next ablate our negative-mining and group-construction pipeline (Section~\ref{sec:negatives}) at 0.6B Raw across three configurations: (1) \emph{Random Negative ($1{:}1$ Ratio)}, which pairs each query--positive example with a single random negative; (2) \emph{Random Negatives ($1{:}3$ Ratio)}, which expands each group to three random negatives; and (3) \emph{Screened Stratified $1{:}3$ (Ours)}, which excludes retrieval candidates with $\Delta \le 1$ and samples across the \emph{Hard} ($j_0$), \emph{Medium} ($j_{25}$), and \emph{Moderate} ($j_{50}$) negative tiers.

\begin{table}[!htb]
\centering\small
\setlength{\tabcolsep}{5.8pt}
\begin{tabularx}{\linewidth}{@{}>{\raggedright\arraybackslash}Xcccccc@{}}
\toprule
\rowcolor{tablehead}
0.6B Raw Data \& Negative Configuration & CMTEB-R & MMTEB-R & BEIR & MLDR & Code & Avg. \\
\midrule
\multicolumn{7}{@{}l}{\emph{\textbf{Data Composition Ablation}}} \\
Public Annotations Only (w/o Synthetic) & 75.34 & 67.53 & 60.20 & 64.52 & 76.26 & 68.77 \\
\rowcolor{tableresult}
\textbf{+ Curated Synthetic Data (Full 2.75M, Ours)} & \textbf{75.76} & \textbf{69.90} & \textbf{60.61} & \textbf{68.34} & \textbf{77.53} & \textbf{70.43} \\
\midrule
\multicolumn{7}{@{}l}{\emph{\textbf{Negative Mining \& Group Construction Ablation}}} \\
Random Negative ($1{:}1$ Ratio) & 71.50 & 66.90 & 55.49 & 66.08 & 75.58 & 67.11 \\
Random Negatives ($1{:}3$ Ratio) & 73.39 & 68.71 & 57.71 & 66.25 & 76.13 & 68.44 \\
\rowcolor{tableresult}
\textbf{Screened Stratified Negatives ($1{:}3$ Ratio, Ours)} & \textbf{75.76} & \textbf{69.90} & \textbf{60.61} & \textbf{68.34} & \textbf{77.53} & \textbf{70.43} \\
\bottomrule
\end{tabularx}
\caption{Ablation of synthetic data expansion and negative-mining strategies on unmerged 0.6B Raw checkpoints evaluated on Qwen Top-100 candidates (retrieved by Qwen3-Embedding-0.6B).}
\label{tab:ablation-synthesis-neg}
\end{table}

As shown in the bottom block of Table~\ref{tab:ablation-synthesis-neg}, expanding from a single random negative ($1{:}1$) to three random negatives ($1{:}3$) raises the five-panel average from 67.11 to 68.44 ($+1.33$), confirming the benefit of wider in-group comparisons. Replacing random negatives with teacher-screened ($\Delta > 1$), difficulty-stratified retrieval candidates (\emph{Screened Stratified $1{:}3$}) delivers an additional $+1.99$ Avg.\ improvement (reaching \textbf{70.43}, or $+3.32$ over Random $1{:}1$), with substantial gains across every panel ($+2.90$ on BEIR, $+2.37$ on CMTEB-R, $+2.09$ on MLDR, and $+1.40$ on Code).

\subsection{Effect of Ranking Distillation Objectives}
\label{sec:ablation-distill}
To test the properties analyzed in Section~\ref{sec:hinge}, we evaluate four training objectives on 0.6B Raw under matched conditions (identical initialization, 2.75M group order, 43,003 optimizer steps, and seed 42):
\begin{itemize}
 \item \textbf{SFT Only ($\lambda=0$):} Pointwise token cross-entropy on \texttt{yes}/\texttt{no} without pairwise teacher distillation.
 \item \textbf{Capped Gap MSE:} Joint SFT plus squared-error regression $(m_{ij}-u_{ij})^2$ on the capped teacher margins ($m_{\max}=4$) across all eligible pairs ($\lambda=0.1$).
 \item \textbf{P--N Hinge Only:} Joint SFT plus minimum-gap hinge loss $[m_{ij}-u_{ij}]_+$ restricted to the three positive--negative pairs per group, omitting negative--negative comparisons ($\lambda=0.1$).
 \item \textbf{Full Hinge (Ours):} Joint SFT plus minimum-gap hinge loss across all six eligible positive--negative and negative--negative pairs (Equation~\ref{eq:totalloss}, $\lambda=0.1$).
\end{itemize}

\begin{figure}[t!]
 \centering
 \includegraphics[width=\linewidth]{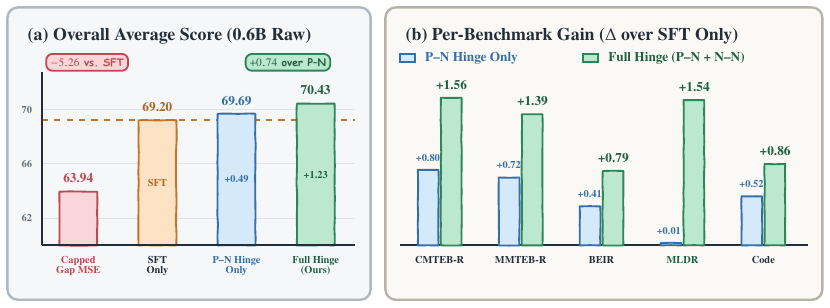}
 \caption{\textbf{Controlled 0.6B Raw comparison across ranking distillation objectives on Qwen Top-100.} \textbf{(a)} Overall Avg.\ scores: Capped Gap MSE degrades performance by $-5.26$ points below SFT Only, whereas P--N Hinge ($+0.49$) and Full Hinge ($+1.23$) yield consistent gains. \textbf{(b)} Per-benchmark relative gain ($\Delta$ over SFT Only), showing how negative--negative (N--N) comparisons unlock a $+1.54$ gain on long-document retrieval (MLDR).}
 \label{fig:distillation-ablation}
\end{figure}

Figure~\ref{fig:distillation-ablation} summarizes the benchmark results across all five evaluation panels:

\paragraph{Why Capped Gap MSE Collapses ($-5.26$ Avg.\ vs.\ SFT Only).}
Replacing the one-sided hinge penalty with squared-error regression (\emph{Capped Gap MSE}) causes a sharp drop across every panel, reducing Avg.\ from 69.20 to 63.94 ($-6.60$ on CMTEB-R, $-6.30$ on BEIR, and $-6.00$ on MLDR). This drop reflects two conflicts analyzed in Figure~\ref{fig:marginshape}: \emph{(i) over-separation penalty}, where SFT naturally drives student logit gaps $u_{ij}$ above $m_{\max}=4$ on clear pairs, causing the symmetric $(m_{ij}-u_{ij})^2$ term to pull down positive scores and push up negative scores to compress the gap back toward $4$; and \emph{(ii) path-additivity conflict}, where independently capped pairwise targets ($m_{ab}+m_{bc}>m_{ac}$) create mutually inconsistent equality constraints across the six pairs in a group (Appendix~\ref{app:margin-consistency}).

\paragraph{One-Sided Lower Bounds and Negative--Negative Contrasts ($+1.23$ Avg.\ over SFT Only).}
Switching from squared error to our one-sided hinge objective removes both conflicts: once $u_{ij}\ge m_{ij}$, the pairwise gradient becomes zero, allowing SFT to calibrate token probabilities without penalty. Restricting the hinge loss to positive--negative pairs (\emph{P--N Hinge Only}) raises Avg.\ to 69.69 ($+0.49$ over \emph{SFT Only}). Adding the three negative--negative pairs in \emph{Full Hinge} lifts Avg.\ to \textbf{70.43} ($+1.23$ over \emph{SFT Only} and $+0.74$ over \emph{P--N Hinge Only}), accounting for \textbf{60\% of the total distillation gain} and driving a $+1.54$ jump on long-document retrieval (MLDR, from 66.80 to 68.34).

\begin{table}[!htb]
\centering\small
\setlength{\tabcolsep}{7.5pt}
\begin{tabularx}{\linewidth}{@{}>{\raggedright\arraybackslash}Xccc@{}}
\toprule
\rowcolor{tablehead}
Group-Level Diagnostic Metric (1,000 Held-Out $1{:}3$ Groups) & SFT Only & Full Hinge (Ours) & Gain \\
\midrule
Positive $>$ Negative Pairwise Accuracy (3,000 P--N Pairs) & 99.7\% & 99.7\% & $+0.0\text{ pp}$ \\
Hard $>$ Medium Ordering Agreement ($j_0 > j_{25}$) & 71.6\% & \textbf{90.5\%} & \textcolor{reportaccent}{\bfseries +18.9 pp} \\
Medium $>$ Moderate Ordering Agreement ($j_{25} > j_{50}$) & 58.7\% & \textbf{78.1\%} & \textcolor{reportaccent}{\bfseries +19.4 pp} \\
\rowcolor{tableresult}
\textbf{Complete Negative Tier Order ($\text{Hard} > \text{Medium} > \text{Moderate}$)} & \textbf{36.4\%} & \textbf{69.7\%} & \textcolor{reportaccent}{\bfseries +33.3 pp} \\
\midrule
Within-Query Negative Score Span ($s_{\mathrm{hard}} - s_{\mathrm{mod}}$, Mean / Median) & $1.95$ / $1.61$ & \textbf{6.29} / \textbf{6.18} & \textcolor{reportaccent}{\bfseries +4.34 (3.2$\times$)} \\
\bottomrule
\end{tabularx}
\caption{Within-group ordering diagnostic across 1,000 held-out $1{:}3$ groups on unmerged 0.6B Raw checkpoints. While SFT Only already separates positives from negatives (99.7\%), Full Hinge expands the within-query negative score span by $3.2\times$ and nearly doubles complete three-tier negative ordering from 36.4\% to 69.7\%.}
\label{tab:score-diagnostic}
\end{table}

\paragraph{Diagnostic: Why Negative--Negative Supervision Matters.}
Table~\ref{tab:score-diagnostic} explains this gain by comparing the predictions of \emph{SFT Only} and \emph{Full Hinge} across 1,000 held-out stratified $1{:}3$ groups. Because binary labels already supervise \texttt{yes} versus \texttt{no}, both models achieve 99.7\% pairwise accuracy on positive--negative pairs. However, because all three negatives share the same \texttt{no} label under SFT, \emph{SFT Only} compresses them into a narrow $1.95$-logit span and orders all three tiers correctly ($\text{Hard} > \text{Medium} > \text{Moderate}$) in only \textbf{36.4\%} of groups. \emph{Full Hinge} widens the within-query negative score span by $3.2\times$ (to $6.29$ logits) and lifts complete three-tier negative ordering to \textbf{69.7\% ($+33.3\text{ pp}$)}, directly improving fine-grained ranking among competing non-relevant candidates.

\FloatBarrier
\subsection{Effect of Model Merging}
\label{sec:interpolation-study}
\begin{figure}[!htb]
 \centering
 \includegraphics[width=\linewidth]{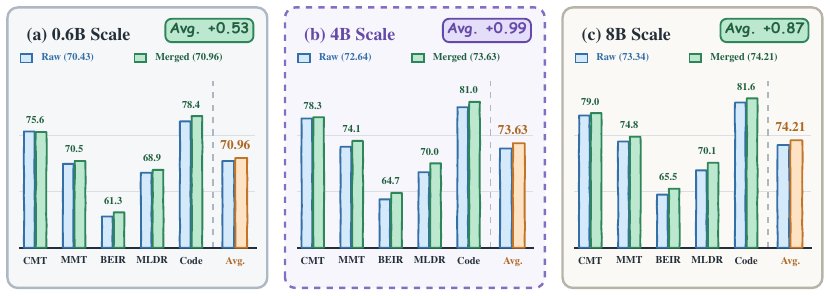}
 \caption{\textbf{Effect of model merging across parameter scales on Qwen Top-100.} \textbf{(a)} At 0.6B, merging lifts Avg.\ from 70.43 to 70.96 ($+0.53$). \textbf{(b)} At 4B, merging lifts Avg.\ from 72.64 to 73.63 ($+0.99$), and \textbf{(c)} at 8B from 73.34 to 74.21 ($+0.87$).}
 \label{fig:merging-ablation}
\end{figure}

Figure~\ref{fig:merging-ablation} compares checkpoints before and after model merging (Section~\ref{sec:merging}). Merging improves Avg.\ across all three model sizes: $+0.53$ points at 0.6B (70.43 $\rightarrow$ 70.96), $+0.99$ points at 4B (72.64 $\rightarrow$ 73.63), and $+0.87$ points at 8B (73.34 $\rightarrow$ 74.21). Across scales, merging consistently strengthens long-document retrieval (MLDR $+0.60$ at 0.6B, $+1.59$ at 4B, and $+1.34$ at 8B) while preserving or improving performance across Chinese, multilingual, English, and code retrieval.

\subsection{Generalization across Candidate Generators}
\label{sec:cross-retriever}
To verify that \model{}'s gains do not depend on the Qwen3-Embedding-0.6B retriever used in our main evaluation, we evaluate all three model scales (0.6B, 4B, and 8B) on Top-100 candidate pools retrieved by Microsoft's Harrier-OSS-v1-0.6B (\textbf{Harrier Top-100})~\citep{microsoft2026harrier} under its native task-description protocol. \mbox{Table~\ref{tab:cross-pool}} compares the \model{} family against the corresponding Qwen3-Reranker baselines across all five benchmark panels.

\begin{table}[!htb]
\centering\small
\setlength{\tabcolsep}{7.2pt}
\begin{tabularx}{\linewidth}{@{}Xcccccc@{}}
\toprule
\rowcolor{tablehead}
Model & CMTEB-R & MMTEB-R & BEIR & MLDR & Code & Avg. \\
\midrule
Qwen3-Reranker-0.6B & 71.16 & 67.55 & 60.30 & 63.25 & 72.90 & 67.03 \\
\rowcolor{tableresult}
\textbf{\model{}-0.6B} & \textbf{75.41} & \textbf{74.81} & \textbf{61.68} & \textbf{65.29} & \textbf{78.15} & \textbf{71.07}~\gain{+4.03} \\
\midrule
Qwen3-Reranker-4B & 75.74 & 73.18 & 64.24 & \textbf{66.01} & 79.92 & 71.82 \\
\rowcolor{tableresult}
\textbf{\model{}-4B} & \textbf{78.11} & \textbf{81.82} & \textbf{65.33} & 65.98 & \textbf{80.60} & \textbf{74.37}~\gain{+2.55} \\
\midrule
Qwen3-Reranker-8B & 77.39 & 72.87 & 65.37 & 65.74 & 80.13 & 72.30 \\
\rowcolor{tableresult}
\textbf{\model{}-8B} & \textbf{78.76} & \textbf{81.21} & \textbf{66.02} & \textbf{66.24} & \textbf{81.15} & \textbf{74.68}~\gain{+2.37} \\
\bottomrule
\end{tabularx}
\caption{Cross-retriever generalization (nDCG@10 $\times 100$) across parameter scales on the shared \textbf{Harrier Top-100} candidate pool (top-100 candidates per query retrieved by Harrier-OSS-v1-0.6B under its native task-description protocol). \textbf{Avg.}\ denotes the equal-weight average across all five panels, and parenthesized gains are computed relative to the same-sized Qwen3-Reranker baseline prior to rounding.}
\label{tab:cross-pool}
\end{table}

As shown in Table~\ref{tab:cross-pool}, \model{} maintains a consistent lead over Qwen3-Reranker across all three parameter scales on Harrier Top-100, improving the five-panel Avg.\ by $+4.03$ points at 0.6B (67.03 $\rightarrow$ 71.07), $+2.55$ points at 4B (71.82 $\rightarrow$ 74.37), and $+2.37$ points at 8B (72.30 $\rightarrow$ 74.68). Notably, \model{}-4B (74.37) again surpasses the Qwen3-Reranker-8B teacher (72.30) by 2.07 points, confirming that our data synthesis and preference distillation recipe generalizes robustly across distinct first-stage candidate generators.

\FloatBarrier
\section{Related Work}
\label{sec:related}
\paragraph{Foundation Models for Reranking.}
Cross-encoders model query--document interactions to estimate relevance~\citep{nogueira2019passage}. Recent rerankers build on foundation models with stronger language understanding and instruction following. Qwen3-Reranker uses instruction-aware pointwise scoring, supervised fine-tuning, and model merging across multiple model sizes~\citep{zhang2025qwen3embedding}. ERank combines supervised learning with reinforcement learning~\citep{cai2025erank}, KaLM-Reranker decouples passage encoding from query--passage cross-attention~\citep{zhao2026kalm}, and Jina Reranker v3 and v3.5 introduce listwise document interaction, hybrid attention, and self-distillation~\citep{wang2025jina,nasika2026jina}. Our work retains the dense pointwise formulation and focuses on the data construction and supervision needed for effective pointwise reranking across model scales.

\paragraph{Synthetic Ranking Supervision.}
InPars-v2 uses language models to generate queries and a reranker to select synthetic pairs~\citep{jeronymo2023inparsv2}; Promptagator studies task-specific query generation for retrieval training~\citep{dai2022promptagator}. Querit-Reranker combines document-conditioned synthetic queries, teacher supervision, and model merging for multilingual distribution adaptation~\citep{zhong2026querit}. These approaches establish synthetic queries as a useful means of adapting language models to relevance tasks. We combine basic and targeted synthesis with quality selection, semantic deduplication, and coverage allocation, then construct natural-negative groups. Teacher-based negative screening is related to denoised hard-negative learning in RocketQA and studies of false negatives~\citep{qu2021rocketqa,cai2022hardnegatives}. Our difficulty-stratified selection uses screened candidates at different positions in the teacher ordering to form each $1{:}3$ group.

\paragraph{Ranking Distillation.}
Teacher models provide supervision beyond discrete relevance labels. RankDistil studies the transfer of ranking information~\citep{reddi2021rankdistil}, and recent reranker experiments examine the advantages of stronger teachers for smaller students~\citep{xu2025distillation}. Margin-MSE distills relative score differences~\citep{hofstatter2020marginmse}; Querit-Reranker also uses teacher-dependent margins in a pairwise hinge objective~\citep{zhong2026querit}. Our recipe adds a cap on the teacher-derived target gap, excludes near-ties and annotation-conflicting groups from distillation, and retains SFT supervision alongside the ranking loss. These constraints operate over the eligible positive--negative and negative--negative comparisons in each constructed group. We systematically ablate both negative-mining strategies and pairwise distillation objectives on controlled data mixtures~\citep{morand2026crossencoders}.

\Needspace{8\baselineskip}
\section{Conclusion}
We introduced \model{}, a family of dense pointwise text rerankers at the 0.6B, 4B, and 8B parameter scales developed through document-grounded query synthesis and minimum-gap preference distillation. Our data pipeline expands intent diversity and domain coverage while using teacher-guided curation and difficulty-stratified negative mining to organize real candidates into informative $1{:}3$ ranking groups. Jointly optimizing pointwise relevance cross-entropy and bounded minimum-gap hinge constraints transfers both binary relevance decisions and fine-grained relative rankings across cross-scale (0.6B and 4B) and same-scale (8B) models, which are further enhanced through layerwise adaptive model merging.

Across five comprehensive benchmarks covering Chinese, multilingual, English, long-document, and code retrieval, \model{} consistently outperforms comparable open-weights rerankers within every parameter tier: the 0.6B model surpasses 1B--4B active-parameter baselines, the 4B model outperforms the 8B teacher, and the 8B model achieves the strongest overall results across all five panels. Controlled ablations and group-level score diagnostics confirm the complementary contributions of curated synthesis, stratified negative mining, and negative--negative preference distillation, while cross-retriever evaluations verify that these gains generalize robustly across distinct candidate generators.

\paragraph{Summary and Future Directions.}
By coupling document-grounded query synthesis, teacher-screened stratified negative mining, and bounded minimum-gap hinge distillation, \model{} delivers state-of-the-art neural text reranking across diverse languages, domains, document lengths, and programming tasks. Future work will explore extending our synthesis and distillation framework to ultra-lightweight architectures and multimodal retrieval settings.

\FloatBarrier
\begingroup
\fontsize{9.5}{11.6}\selectfont
\setlength{\bibsep}{2.4pt}
\installbiburlcleaner
\let\url\underlinedurl
\bibliographystyle{unsrtnat}
\bibliography{references}
\endgroup
\clearpage
\appendix
\begingroup
\fontsize{10.5}{13}\selectfont
\renewcommand{\small}{\fontsize{9.5}{11.7}\selectfont}
\section{Data Recipe Details}
\label{app:data-recipe}
This appendix follows the four stages in Section~\ref{sec:data}, providing data-source attributions, synthesis prompt specifications, curation settings, and negative-selection formulas.

\subsection{Data Collection and Preparation}
\label{app:data-sources}
Table~\ref{tab:data-sources} summarizes representative public supervision sources and unlabelled document collections used in our pipeline. Public supervision sources supply training-split query--positive pairs, while unlabelled document collections provide raw texts for document-grounded synthesis.

\begin{table}[!htbp]
\centering\small
\renewcommand{\arraystretch}{1.22}
\begin{tabular}{@{}>{\raggedright\arraybackslash}p{0.26\linewidth}>{\raggedright\arraybackslash}p{0.70\linewidth}@{}}
\toprule
\rowcolor{tablehead}
Source Category & Representative Datasets and Collections \\
\midrule
Public Supervision Sources &
MS MARCO~\citep{bajaj2016msmarco},
mMARCO~\citep{bonifacio2021mmarco},
Natural Questions~\citep{kwiatkowski2019naturalquestions},
HotpotQA~\citep{yang2018hotpotqa},
2WikiMultiHopQA~\citep{xanh2020_2wikimultihop},
QReCC~\citep{qrecc},
DuReader~\citep{he2018dureader,qiu2022dureaderretrieval},
T$^2$-Ranking~\citep{xie2023t2ranking},
KuaiSearch~\citep{li2026kuaisearch},
MIRACL~\citep{zhang2023miracl},
Mr.\ TyDi~\citep{mrtydi},
MLDR~\citep{chen2024m3},
Multi-CPR~\citep{Long2022MultiCPRAM},
FEVER~\citep{thorne-etal-2018-fever},
SPECTER~\citep{specter2020cohan},
Stack Exchange and Stack Overflow~\citep{stackexchange2026data},
ESCI~\citep{reddy2022shopping},
cMedQA2~\citep{8548603},
WebMedQA~\citep{he2019applying},
MedQuAD~\citep{BenAbacha-BMC-2019},
KUAKE-QTR~\citep{zhang-etal-2022-cblue},
LeCaRDv2~\citep{li2023lecardv2}, and
CodeSearchNet~\citep{husain2019codesearchnet}. \\
\midrule
Document Collections &
Wikipedia~\citep{wikimedia2026dumps},
FineWeb2~\citep{penedo2025fineweb2pipelinescale},
PMC Open Access~\citep{ncbi2026pmc},
MultiEURLEX~\citep{chalkidis-etal-2021-multieurlex},
and GitHub issues~\citep{openindex2026issues}, alongside multilingual web, news, scientific, and open-source software and database documentation collections. \\
\bottomrule
\end{tabular}
\caption{Representative data sources used for supervised fine-tuning and document-grounded synthesis.}
\label{tab:data-sources}
\end{table}

\paragraph{Source Preparation.}
Source adapters preserve the relative order of headings, paragraphs, lists, and preformatted blocks after stripping page chrome. Structured records use their primary body or content fields, retaining language tags and source provenance. Annotation adapters preserve task instructions and known positive links so that annotated positives are excluded during negative mining. All training pairs and candidate pools undergo strict benchmark decontamination against evaluation splits through exact and near-duplicate filtering. Document-grounded synthesis filters source documents prior to generation and enforces the grounding rules in Appendix~\ref{app:synthesis-example}, teacher verification, and semantic deduplication.

\FloatBarrier
\Needspace{10\baselineskip}
\subsection{Basic Synthesis Prompt and Settings}
\label{app:synthesis-example}
\paragraph{Synthesis Decoding and Verification Settings.}
Document-grounded query synthesis uses an instruction-following language model with temperature $0.7$, top-$p$ $0.95$, top-$k$ $50$, and repetition penalty $1.05$. Both basic and query-type-guided synthesis share the same grounding rules, verification checklist, and JSON schema below (with guided synthesis additionally conditioning on a high-level intent category), and downstream relevance is independently judged by the Qwen3-Reranker-8B teacher.

\paragraph{Complete Prompt Instructions.}
The boxes below present the complete basic-synthesis instructions described in Section~\ref{sec:synthesis}. Key operational invariants are highlighted in bold for readability.

\begin{tcolorbox}[breakable,enhanced,colback=reportpale,colframe=reportaccent!45,boxrule=0.5pt,borderline west={2.8pt}{0pt}{reportaccent},arc=1.2mm,left=3.5mm,right=3.5mm,top=2.2mm,bottom=2.2mm,title={Basic Synthesis: System Instructions},colbacktitle=reportaccentlight,coltitle=reportaccent,fonttitle=\sffamily\bfseries,fontupper=\fontsize{9.4}{11.6}\selectfont]
You are a senior multilingual retrieval-data author. Read the complete immutable Source DOC and either write \textbf{one natural standalone retrieval Query} or skip the document when it cannot honestly support a useful Query.

Choose the \textbf{single most natural substantive information need} that a real user could express when trying to retrieve this DOC for information it contains.

\textbf{Answer-First Workflow:} Read the whole DOC $\rightarrow$ identify one useful complete answer $\rightarrow$ locate the shortest complete supporting text needed for that answer $\rightarrow$ write the Query without revealing the answer $\rightarrow$ replay the final Query against the whole DOC. The Query must be answerable from this DOC alone, in the DOC's language unless requested otherwise. \textbf{Never edit, summarize or replace the Positive DOC.}

Return exactly one compact JSON object and no Markdown. Use literal JSON \texttt{null}, never the string \texttt{"null"}.
\end{tcolorbox}

\begin{tcolorbox}[breakable,enhanced,colback=reportpale,colframe=reportaccent!45,boxrule=0.5pt,borderline west={2.8pt}{0pt}{reportaccent},arc=1.2mm,left=3.5mm,right=3.5mm,top=2.2mm,bottom=2.2mm,title={Basic Synthesis: Authoring Rules \& Verification Checklist},colbacktitle=reportaccentlight,coltitle=reportaccent,fonttitle=\sffamily\bfseries,fontupper=\fontsize{9.4}{11.6}\selectfont]
\textbf{1. Choose One Coherent Real Information Need:}\par
$\bullet$ Asks for a substantive fact, rule, explanation, comparison, finding, procedure, condition or resolution actually supported by the complete DOC.\par
$\bullet$ Sounds like something a user who does not know the answer would search for.\par
$\bullet$ Specific enough to retrieve this DOC, but \textbf{does not copy the answer}, enumerate answer-bearing details, or turn the answer into a leading yes/no question.\par
$\bullet$ May require 1--3 supporting text units when naturally necessary; do not force multiple units, distant spans, ambiguity, or artificial difficulty.

\textbf{2. Reject Document-Aboutness and Metadata Shortcuts:} Do not ask what ``this document'', ``the article'', ``the study'', ``the decision'', a passage, table or source says. Do not ask merely for the title, author, publication date, issue number, identifier, language, repository name or a visible heading unless that metadata is itself the substantive information need. A topical summary such as ``What is this document about?'' is not a valid retrieval Query.

\textbf{3. Preserve Exact Scope and Modality:} Actor, object, population, component, time, jurisdiction, stage, condition, exception, \emph{may/can/must}, \emph{support/refute}, and \emph{cause/association} must not be silently altered. Do not repair the DOC with outside knowledge or prefer a familiar claim over the visible text. The complete answer must be uniquely supported under the final Query across title, body, tables, notes, and conclusion.

\textbf{4. Avoid Copying Surface Questions or Headings:} Before choosing an answer, scan the complete \texttt{source\_doc} for any pre-existing questions in titles, headings, abstracts, or FAQ sections. Do not paraphrase, narrow, broaden, or convert existing surface questions into yes/no form. Choose a substantively different answer proposition grounded in the body prose.

\textbf{5. Byte-Exact Supporting Quote Verification:} Return 1--3 independently necessary supporting text units. Copy every \texttt{supporting\_quote} verbatim from \texttt{source\_doc} after the Query is fixed. It must be \textbf{one unique, contiguous, byte-exact substring}: do not normalize whitespace, punctuation, spelling, Unicode, case or typography; do not join separate spans. Verify literal substring equality and uniqueness against \texttt{source\_doc} before returning the candidate.

\textbf{6. Pre-Commit Silent Checklist:}\par
(1) Natural, standalone Query (no document-aboutness or metadata shortcuts).\par
(2) Reading only the Query does not leak the answer.\par
(3) Answer is fully supported by \texttt{source\_doc} alone with exact scope.\par
(4) Every \texttt{supporting\_quote} passes contiguous byte-exact substring replay and uniqueness.\par
(5) Distinct from any surface question or heading already present in the text.\par
(6) \texttt{source\_doc} remains byte-for-byte unchanged.
\end{tcolorbox}

\begin{tcolorbox}[enhanced,colback=reportpale,colframe=reportaccent!45,boxrule=0.5pt,borderline west={2.8pt}{0pt}{reportaccent},arc=1.2mm,left=3.5mm,right=3.5mm,top=2.2mm,bottom=2.2mm,title={Basic Synthesis: JSON Output Schema},colbacktitle=reportaccentlight,coltitle=reportaccent,fonttitle=\sffamily\bfseries,fontupper=\fontsize{9.2}{11.4}\selectfont]
{\ttfamily\small
\{\par
\ \ "status": "AUTHORED|SKIPPED",\par
\ \ "query": "one natural standalone Query in the source language, empty if skipped",\par
\ \ "answer\_proposition": "one complete source-grounded answer, empty if skipped",\par
\ \ "supporting\_passages": [\{\par
\ \ \ \ "unit\_id": "E1",\par
\ \ \ \ "proposition": "the complete necessary proposition supported by this unit",\par
\ \ \ \ "section\_or\_anchor": "a stable visible anchor",\par
\ \ \ \ "supporting\_quote": "one sufficient unique contiguous byte-exact source\_doc substring"\par
\ \ \}],\par
\ \ "standalone\_check": true, "source\_doc\_answerable\_check": true,\par
\ \ "no\_answer\_leakage\_check": true, "positive\_unchanged\_check": true,\par
\ \ "distinct\_from\_headings\_check": true, "author\_confidence": 0.95\par
\}}
\end{tcolorbox}
\FloatBarrier

\subsection{Quality and Coverage Control}
\label{app:selection-settings}
Section~\ref{sec:qpquality} defines positive eligibility, query-neighbor clustering, and coverage allocation (Equations~\ref{eq:positive-quality-selection}--\ref{eq:coverage-selection}). Semantic deduplication encodes queries with Qwen3-Embedding-0.6B (4,096-token limit, $\ell_2$-normalized). Exact query deduplication strips whitespace, punctuation, and underscores after case folding; document deduplication uses exact content identity.

For each source with $n$ qualified positive documents sorted by character length $\ell_{(1)}\le\cdots\le\ell_{(n)}$, the length cutoffs are $a=\ell_{(\lfloor n/3\rfloor+1)}$ and $b=\ell_{(\lfloor 2n/3\rfloor+1)}$. Documents are assigned to the \emph{short} band if $\ell\le a$, the \emph{long} band if $\ell>a$ and $\ell\ge b$, and the \emph{medium} band otherwise. Domain--language budgets are capped at available supply, and integer seat allocation follows Equation~\ref{eq:coverage-allocation}.

\subsection{Negative Mining and Group Construction}
\label{app:negative-settings}
Natural-negative preparation operates over a stored top-1,024 dense retrieval pool per query, sampling candidates after excluding known positives and near-duplicates and filtering by the teacher safety margin ($\Delta(q,n)>1$). For $m\ge3$ eligible negatives sorted by ascending teacher gap $\Delta(q,n)$, the zero-based target indices corresponding to the Hard (top-ranked, $j_0$), Medium (first-quartile, $j_{25}$), and Moderate (median, $j_{50}$) rank positions are:
\begin{equation*}
 \begin{aligned}
 j_0&=0,\\
 j_{25}&=\max\!\left(1,\left\lfloor0.25(m-1)+0.5\right\rfloor\right),\\
 j_{50}&=\min\!\left(m-1,\max\!\left(j_{25}+1,
       \left\lfloor0.50(m-1)+0.5\right\rfloor\right)\right).
 \end{aligned}
\end{equation*}
The selector visits $(j_0, j_{25}, j_{50})$ sequentially and picks the nearest index with a distinct document identity (preferring lower indices on ties).

\FloatBarrier
\section{Training Configuration}
\label{app:train}
Table~\ref{tab:recipe} summarizes the training hyperparameters for \model{}. The 0.6B, 4B, and 8B models use the same training mixture, optimizer, global batch size, learning-rate schedule, and distillation coefficients, initialized from the corresponding instruction-following Qwen3 backbones.

\begin{table}[!htbp]
\centering
\begin{tabular}{ll}
\toprule
\rowcolor{tablehead}
Hyperparameter / Setting & Configuration \\
\midrule
Training Hardware & 64 NVIDIA H20 GPUs \\
Initialization & Qwen3-0.6B, 4B, and 8B (instruction-following backbones) \\
Training Groups & $\approx 2.75\text{M}$ ($1{:}3$ positive-to-negative ratio) \\
Maximum training length & 4,096 tokens (sequence packing disabled; groups kept intact) \\
Global batch size & 256 examples (64 ranking groups across 64 DP workers) \\
Peak learning rate & $10^{-5}$ (cosine decay with 2\% / 861-step warmup) \\
Optimizer & Fused AdamW ($\beta_1=0.9$, $\beta_2=0.999$, $\varepsilon=10^{-8}$) \\
Weight decay / gradient clip & $0.0$ / $1.0$ \\
Training duration & 1 epoch (43,003 optimizer steps; seed 42) \\
Auxiliary distillation weight $\lambda$ & $0.1$ \\
Target margin cap $m_{\max}$ & $4.0$ \\
Teacher tie threshold $\epsilon$ & $0.2$ \\
Precision & BF16 forward/backward; FP32 score and margin arithmetic \\
\bottomrule
\end{tabular}
\caption{Training hyperparameters for \model{}. The SFT-only baseline uses the identical configuration with $\lambda=0$.}
\label{tab:recipe}
\end{table}

Models are trained with full-parameter updates on 64 NVIDIA H20 GPUs. To safeguard human annotations on public benchmarks, group-level distillation is enabled only when the teacher ranks the positive above all negatives ($E_g=\mathbf{1}[t_p \ge \max_k t_{n_k}]$), and near-tied pairs with $|\delta_{ij}|\le\epsilon$ are masked out via $a_{ij}$.

\FloatBarrier
\section{Evaluation Benchmark Details}
\label{app:eval}
Following standard benchmark compositions~\citep{zhang2025qwen3embedding}, all evaluations use MTEB v1.38.9 with FP16 inference, an 8,192-token content prefix limit, and the panel aggregation hierarchy in Table~\ref{tab:aggregation}. Avg. equally averages the five panel scores (CMTEB-R, MMTEB-R, BEIR, MLDR, and MTEB-Code) prior to display rounding.

\begin{table}[!htb]
\centering\small
\begin{tabular}{lrr>{\raggedright\arraybackslash}p{0.51\linewidth}}
\toprule
\rowcolor{tablehead}
Panel & Tasks & Units & Aggregation Rule \\
\midrule
CMTEB-R & 8 & 8 & Equal mean across the 8 Chinese retrieval tasks. \\
MMTEB-R & 16 & 49 & Mean across language subsets within each task, then equal mean of 16 tasks. \\
BEIR & 15 & 26 & Mean of 12 CQADupstack subforums as one task, then equal mean across 15 tasks. \\
MLDR & 1 & 13 & Equal mean across the 13 test-language subsets. \\
MTEB-Code & 12 & 39 & Mean of language subsets within each task, then equal mean of 12 tasks. \\
\bottomrule
\end{tabular}
\caption{Fixed evaluation panel hierarchy. Every evaluated unit contributes an explicit nDCG@10 score.}
\label{tab:aggregation}
\end{table}

\begingroup\raggedright
\textbf{CMTEB-R.} T2Retrieval, MMarcoRetrieval, DuRetrieval, CovidRetrieval, CmedqaRetrieval, EcomRetrieval, MedicalRetrieval, and VideoRetrieval (development splits).

\textbf{MMTEB-R.} AILAStatutes, ArguAna, CovidRetrieval, HagridRetrieval, LegalBenchCorporateLobbying, LEMBPasskeyRetrieval (\texttt{test\_256}), MIRACLRetrievalHardNegatives (18 languages, dev), SCIDOCS, SpartQA, StackOverflowQA, StatcanDialogueDatasetRetrieval (2 subsets, dev), TempReasonL1, TRECCOVID, TwitterHjerneRetrieval, WikipediaRetrievalMultilingual (16 languages), and WinoGrande.

\textbf{BEIR and MLDR.} BEIR comprises the 12 CQADupstack subforums plus ArguAna, ClimateFEVER, DBPedia, FEVER, FiQA2018, HotpotQA, MSMARCO (dev), NFCorpus, NQ, QuoraRetrieval, SCIDOCS, SciFact, Touche2020, and TRECCOVID. MLDR averages test splits across 13 languages (\texttt{ar, de, en, es, fr, hi, it, ja, ko, pt, ru, th, zh}).

\textbf{MTEB-Code.} AppsRetrieval, COIRCodeSearchNetRetrieval (6 languages), CodeEditSearchRetrieval (13 languages), CodeFeedbackMT, CodeFeedbackST, CodeSearchNetCCRetrieval (6 languages), CodeSearchNetRetrieval (6 languages), CodeTransOceanContest, CodeTransOceanDL, CosQA, StackOverflowQA, and SyntheticText2SQL.
\par\endgroup

\section{Compatibility of Capped Margin Constraints}
\label{app:margin-consistency}
For any valid group $g$, the feasible set of student score vectors satisfying all capped lower-bound constraints is
\begin{equation}
 \mathcal{S}_g=\left\{s\in\mathbb{R}^{4}:\operatorname{sign}(t_i-t_j)(s_i-s_j)\geq\min(|t_i-t_j|,m_{\max})\quad\forall\, a_{ij}=1\right\}.
\end{equation}
Because $|t_i-t_j|\ge\min(|t_i-t_j|,m_{\max})$, setting $s=t$ satisfies every inequality, guaranteeing $\mathcal{S}_g\neq\emptyset$. By contrast, exact equality matching ($s_i-s_j=m_{ij}$) violates path additivity ($u_{ac}=u_{ab}+u_{bc}$) whenever $m_{\max}$ clips the outer gap: for $(t_a,t_b,t_c)=(2,-1,-4)$ and $m_{\max}=4$, the capped targets $(3,3,4)$ satisfy $3+3>4$, making zero-loss regression impossible even for an unconstrained student.

\endgroup
\end{document}